\documentclass[prd,twocolumn,showpacs,amssymb,amsmath]{revtex4}
\usepackage[cp1251]{inputenc}
\usepackage{graphicx}
\begin{document}
\title{Dynamical chiral symmetry breaking in $SU(N_{c})$ gauge
theories with large number of fermion flavors}
\author{O.Gromenko}
\email{Oleksandr.Gromenko @ cern.ch}
\affiliation{Department of Physics Kiev Shevchenko National University,\\
pr.Gluskova 2, Kiev 03022, Ukraine\\
\\Department of Physics Clarkson University\\ 8 clarkson ave. Potsdam, NY 13699-5822 USA}

\date{\today}
\begin{abstract}
In this paper we examine a phase transition in $SU(N_{c})$ gauge
theories governed by the existence of an infrared fixed point of the
renormalization group $\beta$ function. The nonlinear integral
Schwinger-Dyson equation for a mass function of massless fermions is
solved numerically using the exact expression of the running
coupling in two-loop approximation for an $SU(3)$ gauge theory.
Based on the obtained solution of the Schwinger-Dyson equation, the
value of the chiral condensate, $\langle\bar{q}q\rangle$, and the
decay constant, $f_{\pi}$, of bound states (mesons) are calculated
for several values of fermion flavors $N_{f}$. We show that this
kind of phase transition is a transition of finite order.
\end{abstract}
 \maketitle

%\newline \tableofcontents
%\newpage
%\begin{multicol}{2}
\numberwithin{equation}{section}
\section{Introduction}
\par Gauge field theories with a large number of
massless fermions are becoming an attractive topic for theoretical
research. A recent discovery of a phase transition in such theories
\cite{PhaseTransition}, \cite{PhaseTransition1} has led to
additional interest. Depending on the number of fermions, there are
two possible phases. The first phase is a phase with broken chiral
symmetry and confinement which occurs when the number of fermions is
less than some critical value $(N_{f}<N_{f}^{cr})$, where
$N^{cr}_{f}$ is a critical value of the number of fermions. The
second phase, occurs when $(N_{f}>N_{f}^{cr})$, is a phase with
strict chiral symmetry and absence of confinement. This type of
phase transition occurs for example in minimal supersymmetrical QCD
\cite{tekhnicolor}. The dynamics of these two phases is well studied
in the approximation discussed bellow.
\par The reason for the phase transition is the existence of an
infrared fixed point (IFP), with coupling constant
$\alpha_{*}=\alpha(0)$, in two loop approximation for the
renormalization group $\beta$ function. Also the mathematical aspect
of the phase transition becomes clear from the analysis of the
integral Schwinger-Dyson equation (SDE) for the mass function.
Assigning to the running coupling the constant value at the IFP,
$\alpha(Q^{2})\equiv \alpha_{*}$, the SDE turns out to be an
equation for eigenvalues and has only trivial solution with
sufficiently small $\alpha_{*}$. In contrast, if $\alpha_{*}$ is
larger than a critical value $\alpha_{c}=\pi/(3C_{2}(F))$, then the
SDE has nontrivial solutions. The critical value $N^{cr}_{f}$ is
then determined from the relation $\alpha_{*}=
\alpha_{*}(N_{f},N_{c})$, at constant $N_{c}$. However the situation
changes when we use the running coupling. The kernel of the integral
SDE, like the running coupling, is a function of the number of
fermions and colors. And it is unclear what should be considered as
a critical value of $\alpha_{*}$. The most obvious solution would be
the introduction of a certain integral characteristic for the SDE
(as has been done for the symmetric kernel in the theory of integral
equations). We could then consider number of fermions as the free
parameter, which would determine the value of $\alpha_{*}$. \par If
a chiral symmetry is broken then there exist boson degrees of
freedom, which of course arise as Goldstone bosons. We examine the
chiral phase transition by studying order parameters like the quark
condensate, $\langle\bar{q}q\rangle$, and the decay constant,
$f_{\pi}$, of bound states. Examining these quantities near the
point of phase transition may shed light on the nature of the
transition. In this paper we study numerical solution of the SDE for
a mass function using an exact expression for the running coupling.
The existence of both trivial and nontrivial solutions of the SDE
confirms that the phase transition takes place in an $SU(3)$ gauge
theory.\par In section \ref{some properties} we discuss the equation
for the running coupling in two loop approximation. In section
\ref{section Schwinger-Dyson equation} we briefly go over the
conservation of Ward-Takahashi identities and set up the SDE in the
local gauge. Section \ref{sectio numerical calculations} is devoted
to the numerical calculations and discussion.
\section{\label{some properties} Some properties of
the gauge theories with the infrared fixed point} Let's start from
the Lagrangian of the $SU(N_{c})$ gauge field theories. It appears
as follows:
\begin{equation}\label{Lagr}
    \mathcal{L}=\sum^{N_{f}}_{k=1}\bar{\psi_{k}}(i\widehat{D})\psi_{k} -
    \frac{1}{4}F^{a}_{\mu\nu}F^{a\mu\nu},
\end{equation}
where $\psi_{k}$ is a four component spinor of flavor $k$,
$D^{\mu}=\partial^{\mu}-igA^{\mu}_{a}T_{a}$, $T_{a}$ are the
generators of the gauge group and $g$ is the coupling constant. This
Lagrangian is clearly invariant under the global symmetry group
$SU(N_{f})_{L}\times SU(N_{f})_{R}\times U(1)_{L,R}$ because all
fermions are massless. Nonetheless, this symmetry may be broken to
diagonal subgroup $SU(N_{f})_{L+R}\times U(1)_{L,R}$.
\par The next step is an analysis of the equation for the running coupling
in two loop approximation (the first two coefficients are
independent of the renormalization scheme, the higher-order
coefficients are scheme dependent). It takes the form:
\begin{equation}\label{beta1}
    \frac{d\alpha}{d\ln(Q^{2}/\nu)}=-b\alpha^{2}-c\alpha^{3}-...,
\end{equation}
where $\alpha=g^{2}/4\pi$ and according to Ref.\cite{secondorder}
coefficients $b$ and $c$ look as follows:
\begin{align}\label{coeff}
  b &= \frac{1}{12\pi}(11C_{2}(A)-4T_{f}N_{f}),\\
  c &= \frac{1}{16\pi^{2}}\left(\frac{34}{3}C_{2}(A)^{2}-
  \frac{20}{3}C_{2}(A)T_{f}N_{f}-4C_{2}(F)T_{f} N_{f}\right),
\end{align}
The theory is asymptotically free if $b>0$. The IFP exists if $c<0$,
which for $SU(3)$ takes place when $N_{f}>8$. The running coupling
at the IFP takes the value $\alpha_{*}=-b/c$. The fixed point
coupling $\alpha_{*}$ can be made sufficiently small to perform a
calculation by perturbation theory. Certainly we assume that
$0<g\lesssim 1$ in the initial Lagrangian (\ref{Lagr}), which
restricts the variation interval for $N_{f}$. For an $SU(3)$ gauge
theory this variation interval is $10\lesssim N_{f}<33/2$ assuming
an asymptotic freedom of the theory.
\par The equation (\ref{beta1}) can be integrated as follows:
\begin{equation} \label{beta2}
    \frac{1}{\alpha(Q^{2})} + \frac{1}{\alpha_{*}}\ln\left(b+\frac{c}{
    \alpha(Q^{2})}\right)=b\ln\frac{Q^{2}}{\Lambda^{2}},
\end{equation}
where we have introduced the scale
\begin{equation}\label{Lambda}
\Lambda^{2}=\nu^{2}\left(
\frac{b}{\alpha_{s}(\nu^{2})}+c\right)^{c/b}
\exp\left(-\frac{1}{b\alpha(\nu^{2})}\right),
\end{equation}
which has the same physical sense as the dimensional $\Lambda_{QCD}$
parameter in ordinary QCD. For further calculations we consider
$\Lambda$ to be independent of $N_{f}$, for a fixed value of
$N_{c}$. This is a good approximation when we are taking into
consideration the small variation interval of $N_{f}$.
\par Equation (\ref{beta2}) is a transcendental equation for
$\alpha(Q^{2})$. It can be solved analytically using complex Lambert
$W_{k}(z)$ function \cite{LambertW}. Lambert's function satisfies
the transcendental equation $W_{k}(z)\exp(W_{k}(z))=z$, where
$k=0,\pm1,\pm2,...$. We note that there are only two possible real
solutions $k=0,-1$ which take place if the argument
$z\geqslant-1/e$. The Lambert W function has simple asymptotics:
$W_{0}(z\rightarrow 0)\sim z-z^{2}$, $W_{-1}$ diverges when
$z\rightarrow0$, and $W_{k}(z\rightarrow \infty)\sim \ln z + 2\pi ik
- \ln(\ln z + 2\pi ik)$. Then we have:
\begin{gather}\label{Lambert}
  \alpha(Q^{2}) = \alpha_{*}\left(1+W_{i}\left(
  -\frac{(Q^{2}/\Lambda^{2})^{b\alpha_{*}}}{ce}
  \right)\right)^{-1},\\ i = \left(
\begin{array}{cc}
  -1, & c>0 \\
  0,  & c<0
\end{array}
\right).\nonumber
\end{gather}
The case $i=-1$ corresponds to the running coupling in ordinary QCD
in two loop approximation (in the perturbative regime). Further,
only the case $c<0$ will be used for calculations. Based on the
properties of Lambert's function, it is possible to obtain the
asymptotic form of (\ref{Lambert}):
%\begin{equation}
\begin{align}
\label{asimptot1}
  \alpha(Q^{2}\rightarrow 0) &= \alpha_{*}\left(
  1+\frac{(Q^{2}/\Lambda^{2})^{b\alpha_{*}}}{ce}\right),\\
\label{asimptot2}
  \alpha(Q^{2}\rightarrow \infty) &= \frac{1}{b\ln(Q^{2}/\Lambda^{2})}
  \left(1+\frac{1}{b\alpha_{*}}\frac{\ln\ln(Q^{2}/\Lambda^{2})
  }{\ln(Q^{2}/\Lambda^{2})}
  \right).
\end{align}
%\end{equation}
It is important to note that as the number of fermions decreases,
the value of asymptote (\ref{asimptot1}) increases, and the value of
asymptote (\ref{asimptot2}) decreases.
\section{\label{section Schwinger-Dyson equation}Schwinger-Dyson equation for the mass function}
\par Since the initial Lagrangian has only massless fermions, we
must require conservation of the vector and axial vector
Ward-Takahashi (WT) identities. It is essential for the conservation
of the axial WT identity that the running coupling must depend on
the same momentum as the gluon propagator \cite{JainMunczek}. The
use of simple Landau gauge and other local covariant gauges violates
the vector WT identity when the ladder approximation with a bare
vertex is studied, and they are not suitable for this purpose. This
problem can be solved by using a nonlocal gauge which depends on the
momentum:
%\begin{widetext}
\begin{equation}\label{gluon}
    D^{\mu\nu}=-i\left(g^{\mu\nu}-\eta(p)\frac{p^{\mu}p^{\nu}}{p^{2}}
    \right)\frac{1}{p^{2}}.
\end{equation}
%\end{widetext}
This nonlocal gauge was proposed by T.Kugo and M.Mitchard
\cite{Kugo}, where $\eta(p)$:
\begin{equation}\label{Kugo}
    \eta(p)=\frac{2}{p^{2}\alpha}\int^{p}_{0}
    dy(y\alpha(y)-y^{2}\alpha^{'}(y)).
\end{equation}
It is clear that function (\ref{Kugo}) coincides with the Landau
gauge at small and large momenta, i.e.,
$\eta(0)=1,\quad\eta(\infty)=1$.
\par The nonlocal Kugo gauge allows us to write the SDE for the dressing fermion
propagator $S(p)=i/(A(p^{2})\hat{p}-B(p^{2}))$. In the ladder
approximation it is given by:
\begin{widetext}
\begin{equation}\label{SD}
    iS(p)^{-1}=\hat{p}-iC_{2}(F)\int\frac{d^{4}k}{(2\pi)^{4}}
    g^{2}((p-k)^{2})D^{\mu\nu}(p-k)i\gamma^{\mu}S(k)i\gamma^{\nu},
\end{equation}
\end{widetext}
where we use a bare vertex $ig\gamma^{\mu}T^{a}$. The another
advantage is that in the Kugo gauge the fermion wave function
renormalisation constant $A(p^{2})$ is equal to one. Using
(\ref{SD}) we may retrieve the SDE for the dynamical mass function
in Euclidean momentum space:
\begin{widetext}
\begin{equation}\label{SDE}
    B(p_{E}^{2})=C_{2}(F)\int\frac{d^{4}k_{E}}{4\pi^{3}}\alpha((p-k)^{2}_{E})
    \frac{4-\eta((p-k)^{2}_{E})}{(p-k)^{2}_{E}}\frac{B(k_{E}^{2})}{k^{2}_{E} +
    B^{2}(k_{E}^{2})}.
\end{equation}
\end{widetext}
However, the obtained equation is rather complicated and can not be
solved analytically without certain assumptions and approximations.
One of these approximations is to use the constant value at the IFP
for the running coupling, i.e., $\alpha(Q^{2})=\alpha_{*}$. It may
be used in the region of momentum where the running coupling is
slowly changing. Or in other words for
$\alpha_{*}\rightarrow\alpha_{c}$ from above and $N_{f}\rightarrow
N_{c}$ from below. Expression (\ref{Kugo}) in this approximation is
equivalent to the Landau gauge and therefore (\ref{SDE}) becomes:
\begin{equation}\label{SDalpha}
    B(p^{2})=3C_{2}(F)\alpha_{*}\int\frac{d^{4}k}{4\pi^{3}}
    \frac{1}{(p-k)^{2}}\frac{B(k^{2})}{k^{2}+B^{2}(k^{2})}.
\end{equation}
%We reproduce well known result obtained by \cite{B0} to show the
%possibility for angular integration only in this simple case.
Equation (\ref{SDalpha}) can be integrated over angular part:
%\begin{widetext}
\begin{equation}\label{angel}
    \int d^{4}k \frac{1}{(p-k)^{2}} %& =4\pi\int\limits_{0}^{\infty}k^{3}dk
    %\int\limits_{0}^{\pi}d\xi\sin^{2}\xi \frac{1}{p^{2}+k^{2}-2\vert p\vert
    %\vert k \vert \cos\xi}\nonumber \\
    %& = 2\pi^{2}\int\limits_{0}^{\infty}k^{3}d k
    %\left(\frac{\Theta(p^{2}-k^{2})}{p^{2}}+\frac{\Theta(k^{2}-
    %p^{2})}{k^{2}}\right)\nonumber \\
    %&
    = \pi^{2}\int\limits_{0}^{p^{2}}k^{2}d k^{2}
    \frac{1}{p^{2}}+\pi^{2}\int\limits_{p^{2}}^{\infty}k^{2}d
    k^{2}\frac{1}{k^{2}}.
\end{equation}
%\end{widetext}
In general this type of angular integration can not be performed
exactly in the more complicated cases. Substituting (\ref{angel})
into (\ref{SDalpha}), we obtain the final integral equation for the
mass function:
\begin{widetext}
\begin{equation}\label{SDsimple}
    B(p^{2})=\frac{3C_{2}(F)\alpha_{*}}{4\pi}\left(\int\limits_{0}^{p^{2}}dk^{2}
    \frac{k^{2}}{p^{2}}\frac{B(k^{2})}{k^{2}+B^{2}(k^{2})}+\int\limits_{p^{2}}^{
     \Lambda^{2}_{*}} dk^{2}
    \frac{B(k^{2})}{k^{2}+B^{2}(k^{2})}\right),
\end{equation}
\end{widetext}
were we have introduced the cutoff parameter $\Lambda_{*}$. By using
simple differentiation with respect to $p^{2}$, this integral
equation converts to a differential equation:
\begin{equation}\label{SDdif}
    p^{2}B''(p^{2})+2B'(p^{2})+\frac{3\alpha_{*}C_{2}(F)}{4\pi}
    \frac{B(p^{2})}{p^{2}+B^{2}(p^{2})}=0,
\end{equation}
with two boundary conditions:
\begin{equation}
\frac{dp^{2}B(p^{2})}{dp^{2}}\vert_{p^{2}=\Lambda^{2}_{*}}=0, \quad
\lim_{p^{2}\rightarrow 0}p^{4}\frac{dB(p^{2})}{dp^{2}}=0.
\end{equation}
The solution of this equation can be found in terms of a
hypergeometric function which has the following asymptotic form at
small momentum:
\begin{equation}\label{miran}
B(0)\approx \Lambda_{*}
exp\left(-\frac{C}{\sqrt{\alpha_{*}/\alpha_{c}-1}}\right),
\end{equation}
The origin of the critical constant  $\alpha_{c}$ now becomes fairly
clear. For recent reviews of the SDE and their application see for
example Refs. \cite{solve}, \cite{rev}.
\section{\label{sectio numerical calculations} Numerical calculations and results}
In this section we discuss the method which was used to solve the
SDE (\ref{SDE}). First of all we note that it is impossible to
perform the angular integration analytically in this case. To solve
the nonlinear SDE, we use a simple quadrature method. All
calculations were performed using Mathematica software and consisted
of the following steps. We set up a square lattice $(p_{i},k_{j})$
where both $p_{i}$ and $k_{j}$ pass the number of discrete values
from lower boundary $q_{0}$ to upper boundary $\Lambda_{*}$ and then
carry out numerical integration using these lattice sites. Next, we
replace the integral in (\ref{SDE}) by quadrature sum of the
rectangles. As a result we obtain a system of nonlinear equations,
where unknown variables serve as the values of the unknown mass
function at the lattice sites:
\begin{equation}\label{grid}
B(p_{i})=\frac{C_{2}(F)}{\pi^{2}}\sum_{k_{j}=q_{0}}^{k_{j}=\Lambda_{*}}
k^{3}_{j}\widetilde{K}(p_{i},k_{j})\frac{B(k_{j})}{k^{2}_{j}+B^{2}(k_{j})}\Delta,
\end{equation}
where $\Delta$ is the lattice step.
\begin{figure}[t]
\includegraphics[width=80mm]{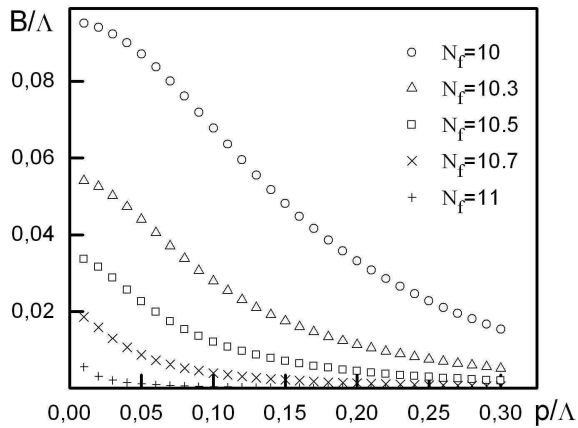}
\caption{\small Numerical solution of the Schwinger-Dyson equation.
  Here $B/\Lambda$ is dimensionless dynamical mass function}\label{b1}
%\hrulefill
\end{figure}
\par Since the unknown function is smooth, it is possible to use a
quadrature formula of rectangles. Also, the same results are
reproduced when we use a more precise quadrature trapezoid formula.
The newly obtained system of nonlinear equations was solved by an
iteration method. We have calculated the mass function for a series
of values of $N_{f}$ near $N^{cr}_{f}$.
\begin{table} \caption{\label{capt1}
Numerical values of the physical quantities obtained from the
numerical solution of the SDE}
\begin{ruledtabular}
\begin{tabular}{lccc}
$N_{f}$ & $B(0)/\Lambda$ & $-\langle\bar{q}q\rangle/\Lambda^{3}\cdot10^{-4}$ & $f_{\pi}/\Lambda$ \\
\hline 10 & 0.0951  & 5.3224 & 0.02791\\
10.3 & 0.0540 & 2.2542 & 0.01845\\
10.5 & 0.0336 & 1.0331 & 0.0126\\
10.7 & 0.0186 & 0.3686 & 0.0077\\
11  & 0.0055  & 0.0473 & 0.0028\\
\end{tabular}
\end{ruledtabular}
\end{table}
\par In Fig.\ref{b1} we illustrate the dependence
$B/\Lambda$ on $p/\Lambda$ for various numbers of fermions. The
behavior of the mass function is as expected: it differs from zero
at small momentum and then go smoothly to zero at large momentum.
Also the quantity $B/\Lambda$ is obviously dimensionless and cannot
depend on the transmutation parameter $\Lambda$. The mass function
$B$ increases when the number of fermions decreases. There is a
similar tendency in (\ref{miran}). The cuttoff $\Lambda_{*}$ is
approximately equal to $0.3\Lambda$. In this area the approximation
$\alpha(Q^{2})\sim\alpha_{*}$ is useful and it is interesting to
compare $B(0)$ with (\ref{miran}). We will do this below.
\par For better understanding of the physical nature of the phase
transition, it is also useful to calculate other physical
quantities. One of them is the value of the vacuum condensate, which
can be easily found:
\begin{equation}\label{condensat}
\langle\bar{q}q\rangle= - \lim_{x\rightarrow +0} tr
S(x,0)=-\frac{N_{c} N_{F}}{8\pi^{2}}\int^{\Lambda_{*}}_{0
}dp_{E}\frac{p^{3}_{E}B(p_{E})}{ p^{2}_{E}+B^{2}(p_{E})},
\end{equation}
and the value of meson decay constant (it is described by well known
Pagels-Stokar formula)
\begin{equation}
\label{PagStokar} f^{2}_{\pi}=\frac{N_{c}
N_{F}}{8\pi^{2}}\int^{\Lambda_{*}}_{0
}dp\frac{p^{3}B(p)}{(p^{2}+B^{2}(p))^{2}} \left(
B(p)-\frac{p^{2}}{2}\frac{B(p)}{dp^{2}}\right)
\end{equation}
These values have also been calculated using quadrature formulas and
are illustrated in Table \ref{capt1}.
\par Let us analyze the quantities $B(0)$, $\langle\bar{q}q\rangle$,
$f_{\pi}$ in more detail. These quantities are continuous functions
near the critical point. This fact confirms that the phase
transition is a transition of second or higher order, possibly  of
infinite order. For this reason, the quark condensate, decay
constant and $B(0)$ may be fitted by polynomial functions near the
critical point.
\begin{align}\label{int}
    B(0) & \sim\left(N^{cr}_{f}-N_{f}\right)^{\alpha},\\
    \langle\bar{q}q\rangle & \sim\left(N^{cr}_{f}-N_{f}\right)^{\beta},\\
    f & \sim\left(N^{cr}_{f}-N_{f}\right)^{\rho}.
\end{align}
If we find that the critical exponents $\alpha$, $\beta$, $\rho$ are
small real numbers, it will confirm that the phase transition is of
finite order. Note, that Eg. (\ref{miran}) describes a phase
transition of infinite order as oppose to finite. The least squares
fitting of the curves gives:
\begin{align}
B(0): & \quad\alpha=2.45, \quad N^{cr}_{f}=11.43;\\
\langle\bar{q}q\rangle: &\quad \beta=2.93 , \quad N^{cr}_{f}=11.21;\\
f: & \quad\rho=1.69 , \quad N^{cr}_{f}=11.33.
\end{align}
We indeed find that the critical exponents are small positive
numbers. The value of the critical number of fermions is described
by the well known formula:
$$N^{cr}_{f}=N_{c}
\left(\frac{100N^{2}_{c}-66}{25N^{2}_{c}-15}\right),$$ obtained from
the condition $\alpha_{*}$ and $\alpha_{c}$ \cite{PhaseTransition}.
In case $N_{c}=3$, the critical number of fermions is
$N^{cr}_{f}=11.9$.
\section{Concluding remarks}
In this paper we have shown numerically that the SDE for a fermion
propagator with an exact expression for the running coupling has
nontrivial solution $B(p)$. Based on the obtained solution, the
value of chiral condensate and the decay constant of pseudoscalar
bosons were calculated. These physical quantities are continuous
functions near the critical point. Detailed analysis of $B(0)$ near
the critical point shows that the phase transition is a transition
of finite order.
\section*{Acknowledgments}
Author would acknowledge V.P.Gusynin for many helpful discussions
and useful notations. We also thank S.I.Vilchinsky for his support
and the warm hospitality at the Kiev Taras Shevchenko University.
Also we thank Daniel T. Robb for careful reading and correction
grammar mistakes.

%\end{twocolumn}

\begin{thebibliography}{00}
\bibitem{PhaseTransition}
T.~Appelquist, A.~Ratnaweera, J.~Terning and L.~C.~R.~Wijewardhana,
Phys. Rev. \textbf{D58}, 105017 (1998).
\bibitem{PhaseTransition1}
T.~Appelquist, J.~Terning and L.~C.~R.~Wijewardhana, Phys.~Rev.~
Lett. \textbf{77}, 1214 (1996). V.~A.~Miransky, K.~Yamawaki, Phys.~
Rev. \textbf{D55}, 105017 (1996).
\bibitem{tekhnicolor}
N.~Seiberg, Phys.~Rev. \textbf{D 49}, 6857 (1994).
\bibitem{secondorder}
W.~E.~Casswell, Phys.~Rev.~Lett. \textbf{33}, 224 (1974).
D.~T.~R.~Jones, Nucl.~Phys. \textbf{B75}, 730 (1974).
\bibitem{LambertW}
R.~M.~Corless, G.~H.~Gonnet, D.~E.~G.~Hare, D.~J.~Jeffrey and D.~E.~
Knuth, Adv.~Comput.~Math. {\textbf{5}},329 (1996).
V.~Elias,G.~McKeon Canadian Journal of Phys. \textbf{84}, 2006.
\bibitem{solve}
P.~I.~Fomin, V.~P.~Gusynin, V.~A.~Miransky and Yu.~A.~Sitenko, Riv.
Nuovo Cimento \textbf{6},1 (1983).
\bibitem{rev} C.~D.~Roberts and A.~G.~Williams, Prog. Part. Nucl. Phys. \textbf{33},477 (1994).
V.~A.~Miransky, Dynamical Symmetry Breaking in Quantum Field
Theories, (World Scientific, Singapore 1993).
\bibitem{JainMunczek}
H.~J.~Munczek and P.~Jain, Phys.~Rev. D \textbf{44}, 1873 (1991).
\bibitem{Kugo}
T.~Kugo and M.~G.~Mitchard, Phys.~Lett. B \textbf{282}, 162-170
(1992).
\bibitem{B0}
P.~I.~Fomin, V.~P.~Gusynin and V.~A.~Miransky, Phys.Lett.
\textbf{78B}, 136 (1978). V.~A.~Miransky, Nuovo Cimento A
\textbf{90}, 149 (1985); Int. J. Mod. Phys. A \textbf{8}, 135
(1993).
\bibitem{fdecay}
H.~Pagels and S.~Stokar, Phys.Rev. \textbf{D20}, 2974 (1979).
H.~Pagels and S.~Stokar, Phys.Rev. \textbf{D22}, 2876 (1980).
J.~M.~Cornwall, Phys.Rev. \textbf{D22}, 1452 (1980).
\bibitem{Cases}
K.-I.~Akoi, M.~Bando, T.~Kugo and M.~G.~Mitchard, Prog. Theor.Phys.
85 (1991) 355.
\end{thebibliography}
\end{document}